\documentclass[conference]{IEEEtran}
\IEEEoverridecommandlockouts
\usepackage{svg}
\usepackage{soul}
\usepackage{cite}
\usepackage{amsmath,amssymb,amsfonts}
\usepackage{graphicx}
\usepackage{textcomp}
\usepackage{xcolor}
\usepackage{caption}
\usepackage{adjustbox}
\usepackage{subcaption}
\usepackage[left=0.680in, right=0.620in, bottom = 1in, top=0.7in]{geometry}
\usepackage{url,amssymb,threeparttable,multirow,booktabs, tabularx}
\def\BibTeX{{\rm B\kern-.05em{\sc i\kern-.025em b}\kern -.08em
    T\kern-.1667em\lower.7ex\hbox{E}\kern-.125emX}}
\usepackage{algpseudocode} 
\usepackage{pifont}

\begin{document}

\title{Optimal Power Allocation and Time Sharing in Low Rank Multi-carrier Wi-Fi Channels}

\author{
\IEEEauthorblockN{Sagnik Bhattacharya\textsuperscript{1}, Kamyar Rajabalifardi\textsuperscript{1}, Muhammad Ahmed Mohsin\textsuperscript{1}, Rohan Pote\textsuperscript{2}, John M. Cioffi\textsuperscript{1}} 

\IEEEauthorblockA{\textsuperscript{1}Dept. of Electrical Engineering, Stanford University, Stanford, CA, USA}

\IEEEauthorblockA{\textsuperscript{2}Samsung Semiconductors, San Diego, CA, USA}

\IEEEauthorblockA{
Emails: \{sagnikb, kfardi, muahmed, cioffi\}@stanford.edu
, rohan.pote@samsung.com}
}

\maketitle
\vspace{-40pt}

\begin{abstract}
The ever-evolving landscape of distributed wireless systems, e.g. multi-user AR/VR systems, demands high data rates (up to 500 Mbps per user) and low power consumption. With increasing number of participating users, uplink data transmission in the situation where the number of transmitter user antennas exceeds the number of access point (AP) antennas presents a low-rank channel problem. Current Wi-Fi standards using orthogonal multiple access (OMA) fail to address these requirements. Non-orthogonal multiple access (NOMA)-based systems, while outperforming the OMA methods, still fall short of the requirement in low-rank channel uplink transmission, because they adhere to a single decoding order for successive
interference cancelation (SIC). This paper proposes and develops a novel optimal power-subcarrier allocation algorithm to maximize the achieved data rates for this low-rank channel scenario. Additionally, the proposed algorithm implements a novel time-sharing algorithm for simultaneously participating users, which adaptively varies the decoding orders to achieve higher data rates than any single decoding order. Extensive experimental validations demonstrate that the proposed algorithm achieves 39\%, 28\%, and 16\% higher sum data rates than OMA, NOMA, and multi-carrier NOMA baselines respectively, under low-rank channel conditions, under varying SNR values. We further show that the proposed algorithm significantly outperforms the baselines with varying numbers of users or AP antennas, showing the effectiveness of the optimal power allocation and time-sharing.
\end{abstract}


\section{Introduction}

The emergence and meteoric rise in Wi-Fi-based distributed wireless systems for mixed reality (MR) requires simultaneous high data rate ($\geq 500$ Mbps) and low power consumption ($\leq 100$ mW). There is also an increase in the number of users/edge devices simultaneously participating in any such application. 
This paper focuses on the uplink multiple-access channel (MAC), where multiple transmitters (e.g., smartphones, gaming consoles, AR/VR glasses) are sending information to a single receiver (e.g., Wi-Fi access point)\cite{book}. 
With increasing number of participating user devices at a given time, the number of uplink transmitter antennas exceeds the number of access point (AP) antennas, presenting an unavoidable low-rank channel problem. 
Existing OMA power allocation methods, conforming to the latest Wi-Fi standards (802.11b/g/n/ac/be) \cite{ieee80211ax2021}, which employ linear receivers, fail to satisfy the high data rate requirements in these low rank channel scenarios. These linear receiver-based systems, which consider all other users' signals as noise while decoding one user's signal, fail to exploit the crosstalk between user channels. Additionally, the current highest achievable data rates through Wi-Fi lead to prohibitive power consumption levels. This scenario is particularly challenging for devices like augmented reality (AR) glasses, where limited power resources are a critical constraint. 


State-of-the-art NOMA power-allocation methods achieve higher data rates and lower power consumption compared to OMA methods \cite{noma6, tse_viswanath_2005}. The authors in \cite{noma6} analyze NOMA for single-input-single-output (SISO) case, where the optimal decoding order for successive interference cancellation (SIC) is simply given by the channel strength (or channel coefficient) between each user and the AP. However, unlike SISO, there is no natural order given by the users' channels in multi-user MIMO scenarios \cite{noma7}. This leads to several heuristic assumptions on the decoding order for MIMO scenarios, e.g., assuming only two users in the system \cite{noma8}, pairing users in groups of 2 and applying NOMA only inside each cluster \cite{noma9}, etc. These heuristic assumptions lead to data rates and power consumption values which fall short of the current requirements. Moreover, the above papers assume the entire bandwidth is allocated to all users, and a power domain / code domain multiple access only, which also contributes to subpar performance. The authors in \cite{noma11} explore multi-carrier hybrid NOMA (MC-NOMA), where power as well as subcarrier allocation is optimized, for the downlink case. However, they also assume a heuristic decoding order for each user, based on the absolute value of the channel coefficient scalar for that user. This does not hold when the channel is a vector for multiple antennas at the AP and/or each user.


To address the crucial data rate / power requirements, and overcome the previous shortcomings of the previous work, in this paper, we propose and implement the optimal power and subcarrier allocation for uplink multiple access channels, with more than 1 antenna at the AP. This approach eliminates heuristic decoding-order assumptions at the receiver (AP), and instead finds the optimal order. Decoupling the derivations of the optimal power-subcarrier allocation and the optimal decoding order helps maintain convexity for both of these problems, which leads to efficient and optimal solutions for both. This is unlike the non-convex power allocation optimization problems introduced in related work \cite {noma5, noma6, noma7, noma8}, because of the heuristic assumptions on decoding order. The new method uses a novel time-sharing algorithm. This time-sharing allows adaptive sharing of multiple decoding orders to achieve data rates higher than those possible using a single decoding order. The proposed power allocation and optimal decoding order derivation, coupled with adaptive time-sharing, helps the user achieve the required data rates/ power consumption values that the current OMA/MC-NOMA methods cannot achieve. Extensive experimental validation shows that the proposed algorithm achieves 39\%, 28\% and 16\% higher data rates compared to OMA, NOMA, and MC-NOMA baselines, under varying SNR levels respectively. 
%




\section{Methodology}

\begin{figure}
    \centering
    \includegraphics[width=0.4\textwidth, height=3.4cm]{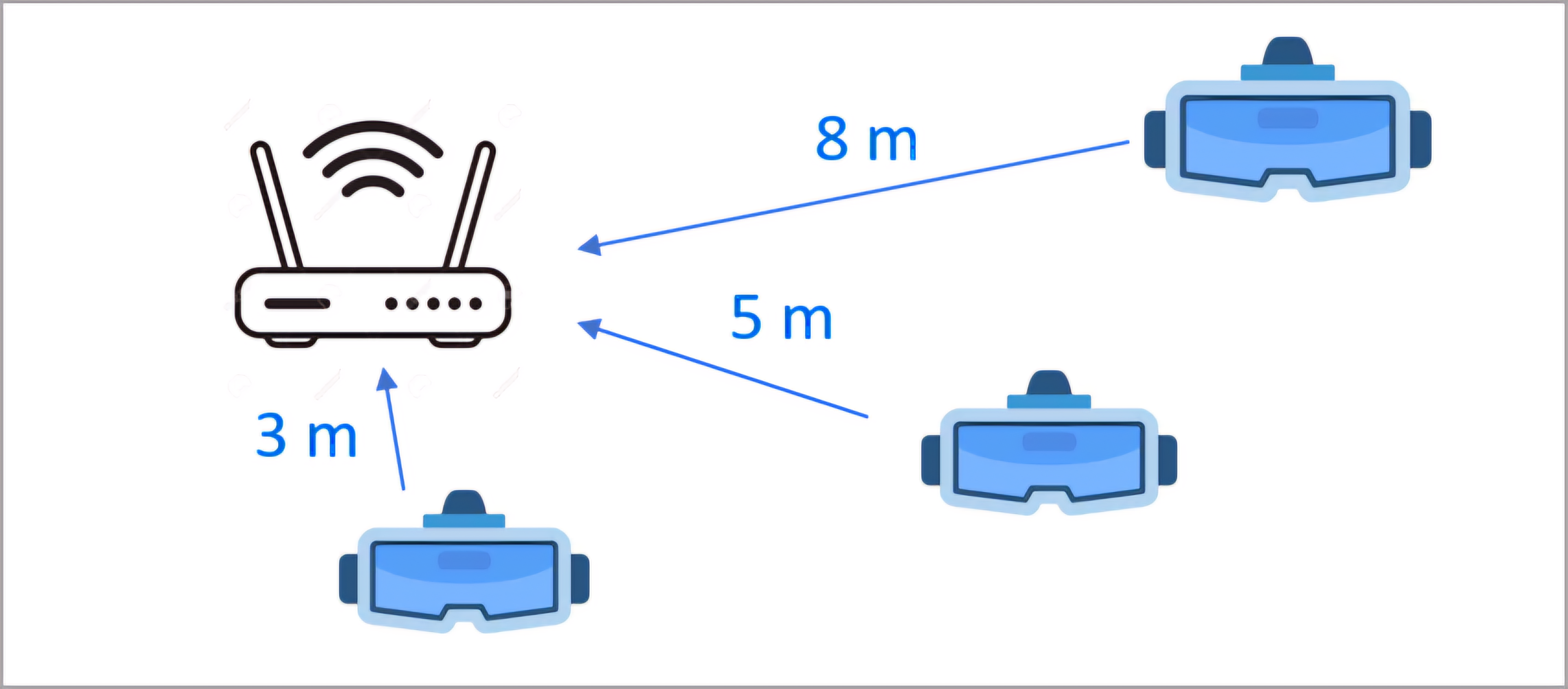}
    \caption{Uplink multiple access channel (MAC)}
    \label{fig:mac}
\end{figure}


\subsection{Background}
Fig~\ref{fig:mac} illustrates the uplink multiple access channel (MAC), where a group of users transmit symbols to an access point (AP). There are $U$ users, where the $u^{th}$ user has $L_{x, u}$ antennas, and transmits symbol vector $\boldsymbol{x}_u \in \mathbb{R}^{L_{x,u}}$ to the AP. The AP has $L_y$ antennas. Therefore, the signal $\boldsymbol{y}$ received at AP is:
\begin{equation} \label{eq:MIMO}
    \boldsymbol{y} = H \:.\: \boldsymbol{x} + \boldsymbol{n}
\end{equation}
where $\boldsymbol{x} \in \mathbb{C}^{L_x}$ concatenates all symbols $\boldsymbol{x}_u$ sent by the users to the receiver, and $\boldsymbol{y} \in \mathbb{C}^{L_y}$ is the received signal. Here $L_x = \sum_{u=1}^{U}L_{x,u}$. Moreover, $H \in \mathbb{C}^{L_y \times L_x}$ denotes the channel matrix, and $\boldsymbol{n} \in \mathbb{C}^{L_y}$ is the additive white Gaussian noise (AWGN) at the receiver. Equation~\eqref{eq:MIMO} can be separated for every subcarrier.
For detection of transmitted symbols at the receiver, we use successive interference cancellation (SIC). For a given user power allocation, a canonical SIC decoding order achieves the information theoretic data rate sum capacity bound with the optimal decoding-order is chosen \cite{book}. Suppose $\boldsymbol{\pi}(.)$ is the decoding order vector, i.e., $\boldsymbol{\pi}(1)$ is the decoding order for user 1, and so forth. Conversely, $\boldsymbol{\pi}^{-1}(.)$ is the argument of $\boldsymbol{\pi}(.)$, where $\boldsymbol{\pi}^{-1}(1)$ indicates the user index that is decoded first. The achieved data rate of user $\boldsymbol{\pi}^{-1}(u)$ is \cite{book}:
\begin{equation}\label{eq:SIC}
    \begin{aligned}
    &b_{\boldsymbol{\pi}^{-1}(u)} = \sum_{i=u}^U b_{\boldsymbol{\pi}^{-1}(i)} - \sum_{i=u+1}^U b_{\boldsymbol{\pi}^{-1}(i)} \\
    &\small{= \operatorname{log}_2 \left(\left|\frac{R_{\boldsymbol{nn}} + \sum_{i=u}^U H_{\boldsymbol{\pi}^{-1}(i)}\cdot R_{\boldsymbol{xx}}(\boldsymbol{\pi}^{-1}(i))\cdot H^*_{\boldsymbol{\pi}^{-1}(i)}}{R_{\boldsymbol{nn}} + \sum_{i=u+1}^U H_{\boldsymbol{\pi}^{-1}(i)}\cdot R_{\boldsymbol{xx}}(\boldsymbol{\pi}^{-1}(i))\cdot H^*_{\boldsymbol{\pi}^{-1}(i)}} \right|\right)}
    \end{aligned}
\end{equation}
where $R_{\boldsymbol{nn}}$ is the noise auto-correlation matrix, and $H_{\boldsymbol{\pi}^{-1}(u)}$ denotes the channel matrix between AP and user $\boldsymbol{\pi}^{-1}(u)$. Equation~\eqref{eq:SIC} follows from the SIC assumption that every user subtracts the interference caused by the already decoded symbols of other users which came before itself in the decoding order. This SIC treats the remaining users' signals as noise. Deriving the optimal decoding order $\boldsymbol{\pi}$ among $U!$ possible orders is non-trivial when it comes to vector channels.


\subsection{Optimal Power and Subcarrier Allocation} \label{Esum}
The optimal power and subcarrier allocation formulates as a primal-dual optimization problem. The primal problem deals with minimization of the users' energy (equivalently power) sum of the users, given the minimum data rates of each. Conversely, the dual problem attempts to maximize
the data rate sum of the users given their maximum energy budget. The weighted energy-sum minimization problem is: 
\begin{equation} \label{eq:energy-sum-minimize}
\begin{aligned}
&\min_{\left\{R_{\boldsymbol{x} \boldsymbol{x}}{(u, n)}\right\}} \quad \sum_{u=1}^U \sum_{n=1}^{N} w_u \cdot \operatorname{trace}\left\{R_{\boldsymbol{x} \boldsymbol{x}}(u, n)\right\} \\
        &\text{C1}: \mathbf{b}=\sum_{n=1}^{N}\left[b_{1, n}, b_{2, n}, \ldots ,b_{U, n}\right]^T \succeq \mathbf{b}_{\min } \succeq \mathbf{0} \\
\\
&\text{C2}: \footnotesize{\sum_{u \in T}\sum_{n=1}^{N} b_{u, n} \leq \operatorname{log}_2\left|\frac{R_{\boldsymbol{nn}} + \left(\sum_{u\in T} H_{u, n} \cdot R_{\boldsymbol{x}\boldsymbol{x}}(u, n) \cdot H_{u, n}^*\right)}{R_{\boldsymbol{nn}}}\right|} \\
&\qquad \forall \: T  \subseteq \{1, 2, \cdots, U\} \\
\\
&\text{C3}: R_{\boldsymbol{x}\boldsymbol{x}}(u,n) \succcurlyeq \mathbf{0} \qquad \forall \: u \in \{1, 2, ..., U\} \\
&\qquad \qquad \qquad \:\qquad \qquad \forall \: n \in \{1, 2, ..., N\}
\end{aligned}
\end{equation}
\vspace{0.02in}
where $R_{\boldsymbol{xx}}(u,n)\in\mathbb{R}^{L_{x,u}\times L_{x,u}}$ is the auto-correlation matrix of $\boldsymbol{x}$ for user $u$ on the $n^{th}$ tone, $w_u$ is a non-negative weight given to $u^{th}$ user's energy in the sum minimization. $b_{u,n}$ is the achieved data rate for the $u^{th}$ user on the $n^{th}$ subcarrier, and $\mathbf{b}_{min}$ is a $U$ dimensional vector representing the minimum required user data rates. C1 represents the minimum data rates constraint, while C2 represents the constraint that for any subset of users, the sum of their data rates should be less than or equal to the information theoretic capacity bound for those users. Finally, C3 denotes that the autocorrelation matrix $R_{\boldsymbol{x}\boldsymbol{x}}(u,n)$ must be a positive semi-definite matrix, i.e  it should be a symmetric matrix and all of its eigenvalues must be non-negative. The non-negative weights $w_u$ reflect the relative importance of minimizing each user's power contribution.

The above optimization problem returns the optimal user power allocations on the subcarriers to achieve the required data rates. This optimization determines the SIC decoding order, and is a convex optimization problem. Thus, decoupling the derivations of the optimal power and subcarrier allocation, and the optimal decoding order, helps maintain convexity and thus achieve efficient and optimal solution to the above problem. This is unlike the non-convex power allocation optimization problems introduced in related work \cite{noma5, noma6, noma7, noma8}, resulting from their heuristic assumptions of decoding order. Given the optimal power allocations on all subcarriers, we now show how to derive the optimal decoding order.

{\color{black}{\subsection{Optimal Decoding Order Derivation}

To derive the optimal decoding order, we first focus on the dual problem of Equation~\eqref{eq:energy-sum-minimize}, which is the data rate sum maximization problem, given maximum energy budget constraints. The dual data rate maximization problem is formulated as:
\begin{equation}
\begin{aligned}
&\max_{\left\{R_{\boldsymbol{x} \boldsymbol{x}}{(u, n)}\right\}} \quad \sum_{u=1}^U \theta_u \cdot\left\{\sum_{n=0}^{N} b_{u, n}\right\} \\ 
& \text{C1}: \mathcal{E}=\sum_{n=0}^{\bar{N}}\left[\mathcal{E}_{1, n}, \mathcal{E}_{2, n}, \ldots, \mathcal{E}_{U, n}\right]^T \preceq \boldsymbol{\mathcal{E}}_{\max } \;\; \\ \\
&\text{C2}: \footnotesize{\sum_{u \in T}\sum_{n=1}^{N} b_{u, n} \leq \operatorname{log}_2\left|\frac{R_{\boldsymbol{nn}} + \left(\sum_{u\in T} H_{u, n} \cdot R_{\boldsymbol{x}\boldsymbol{x}}(u, n) \cdot H_{u, n}^*\right)}{R_{\boldsymbol{nn}}}\right|} \\
&\qquad {\forall \: T  \subseteq \{1, 2, \cdots, U\}} \\ \\
&\text{C3}: R_{\boldsymbol{x}\boldsymbol{x}}(u,n) \succcurlyeq \mathbf{0} \qquad \forall \: u \in \{1, 2, ..., U\} \\
&\qquad \qquad \qquad \:\qquad \qquad \forall \: n \in \{1, 2, ..., N\}
\end{aligned}
\end{equation}

where $\mathcal{E}_{u, n} = trace\{R_{xx}(u,n)\}$ is the energy allocated to the $u^{th}$ user on the $n^{th}$ subcarrier, and $\theta_{u}$ represents the non-negative weights assigned to the user data rates. It is to be noted that $\theta_{u}$ (for all $u \in \{1,2,\cdots,U\}$) are the dual variables corresponding to the minimum required data rate constraints C1 in the primal problem Equation~\eqref{eq:energy-sum-minimize}. It is proven in \cite{book} that the maximum weighted data rate sum must be achievable. Furthermore, it is also proven that, for this optimal data rate sum, the decoding order $\mathbf{\pi}$ must satisfy the following inequality in terms of the dual variables $\theta_{u}$.

\begin{equation}
\label{eq:decoding_order}
    \theta_{\mathbf{\pi}^{-1}(U)} \geq \theta_{\mathbf{\pi}^{-1}(U-1)} \geq ... \geq \theta_{\mathbf{\pi}^{-1}(1)}
\end{equation}

Detailed proof of the above inequality is given in Theorem 5.4.1 of \cite{book}. Hence, it is proved that the required optimal decoding order is given by the order of the dual variables corresponding to the minimum required data rate in the energy sum minimization problem. Since we solve this convex optimization problem, given in Equation~\eqref{eq:energy-sum-minimize}, using a primal-dual approach, we automatically obtain the dual variables $\theta$, and hence the optimal decoding order.

\subsection{Time Sharing}{
A strict ordering of the $\theta_u$ for all $u \in \{1, 2, ..., U\}$ ensures a unique decoding order by which the required data rates can be achieved. However, Equation~\eqref{eq:decoding_order} does not ensure strict inequalities. In cases where one or more of the $\theta_u$ become equal, there is no unique decoding order which can achieve the required data rates~\cite{book}. The related work on NOMA-based methods, which assume a heuristic-based single decoding order, would fail to achieve the required data rates in this case. To get around this, we propose a novel time-sharing solution, which combines multiple decoding orders to achieve the required data rates, in such cases where any individual decoding order fails. The proposed time-sharing algorithm tackles scenarios with identical $\theta_u$ values, by first identifying users with matching $\theta_u$ values and grouping them into  clusters. Inside each cluster, all permutations of the users belonging in it are considered as potential decoding orders. 

Externally, each cluster behaves as a single compound user. The compound users and the remaining single users, which are not part of any cluster, participate in a strict inequality based decoding order. We consider all permutations of users inside every cluster and thus end up with several potential decoding orders. We assume there are $num\_ord$ such candidate decoding orders. It can be proven that all of these candidate decoding orders have the same power allocation requirement\cite{book}. This fact helps us decouple the optimal power allocation in Equation~\eqref{eq:energy-sum-minimize} and the optimal decoding order derivation. The time sharing algorithm next identifies the data rates achieved with each of the candidate decoding orders. This is simply done by taking the optimal power allocation given by Equation~\eqref{eq:energy-sum-minimize}, and applying SIC at the AP with the given candidate decoding order. We assume that the achieved user data rates for the $i^th$ candidate decoding order is given by $s_i$. Finally, the algorithm derives a linear combination of these achievable data rates through linear programming such that the weighted sum is equal to the required data rates. The linear combination coefficients, which are also called the time sharing weights $t\_w$, serve as the time fractions for which the corresponding candidate decoding order will be used. Hence, the proposed algorithm returns the fractions for which several decoding orders must be used to achieve the required data rates, solving the problem in the case where any single decoding order fails. Mathematically, the linear programming component of the time sharing algorithm can be formulated as:

\begin{align*}
\text{minimize} \quad & \sum_{i=1}^{\text{$num\_ord$}} z_i \\
\text{subject to} \quad & \text{C1: } z_i \in \{0, 1\} \quad \forall i \in \{1, \dots, \text{$num\_ord$}\} \\
& \text{C2: } t\_{w_i} \le z_i \quad \forall i \in \{1, \dots, \text{$num\_ord$}\} \\
& \text{C3: } t\_{w_i} \ge 0 \quad \forall i \in \{1, \dots, \text{$num\_ord$}\} \\
& \text{C4: } \sum_{i=1}^{\text{$num\_ord$}} t\_{w_i} = 1 \\
& \text{C5: } \sum_{i=1}^{\text{$num\_ord$}} s_i \cdot t\_{w_i} = \mathbf{b}_{\min }
\end{align*}

This introduces a binary array $z$ of length $num\_ord$. The minimization objective and constraint C2 together ensure minimization of the number of non-zero time-sharing coefficients, i.e. this minimizes the number of participating decoding orders used for time sharing. This is because, changing decoding orders during time sharing introduces latency that this method reduces. Constraints C3 and C4 ensure that the time sharing coefficients are non negative values that sum to 1, since these represent fractions of time for which the corresponding decoding order is used. Finally, C5 represents the crux of the algorithm, and causes the time-shared average data rates to be equal to the required user data rates. 

An example illustrates the time-sharing scenario in Table~\ref{tab:time-sharing2}. As shown in the table, the design tries to achieve target data rates of 500 Mbps per user for a low rank channel with 2 AP antennas and 3 users with 1 antenna each. The users are at equal distances of 3m each from the AP, and the proposed algorithm determines optimal power-subcarrier allocation as well as time sharing. The transmit-power values required by the 3 users in the no time sharing scenario are \{15.8, 16, 15.4\} dBm respectively, while the values with time sharing are lower at \{15, 14.3, 15\} dBm respectively. As seen in the table, time sharing offers the benefit of varying the decoding order in SIC dynamically for given fractions of time, to achieve the required data rates on average, with lower power.

}

}}

        



\section{Baseline Methods}
\label{sec:baselines}
The proposed algorithm improves upon the three baseline methods. The first is OMA resource allocation \cite{oma}, currently used in the WiFi standards, where separate frequency-time resource blocks are allocated to participating users using OFDMA. The receiver in this case is a Fourier transform followed by a linear equalizer. The second baseline is power domain NOMA \cite{noma5, noma6}, where all subcarriers are allocated to all participating users, and different transmit powers are given to each user. The receiver uses SIC with the heuristic assumption of the decoding order given by the channel strength order \cite{noma5, noma6}. The third baseline is hybrid MC-NOMA \cite{noma11} that assigns power and subcarriers to multiple users, but again makes the sub-optimal decoding-order assumption according to the channel strength. 

\section {Performance Evaluation}




\begin{table}[t!]
    \caption{Experiment Parameters}
    \centering
    \resizebox{7.5cm}{!}{
    \begin{tabular}{|c|c|}
    \hline
      Parameter & Value \\
    \hline
      Number of users $U$ & 2-10 \\
      Number of antennas per user $L_{xu}$ & 1 \\
      Number of antennas at AP $L_y$ & 2-32 \\
      User to AP distance range $[d_{min}, d_{max}]$ & [1m, 10m]\\
      Center frequency & 5 GHz \\
      Channel bandwidth $W$ & 80 MHz \\
      Number of subcarriers $N$ & 64 \\
      SNR & [-10, 50] dB \\
      Noise power spectral density & -174 dBm \\
      Noise power & -94.9 dBm \\
    \hline
    \end{tabular}}
    \label{tab:experiment-parameters}
\end{table} 

\begin{table}[t!]
\centering
\caption{Time Sharing Transmit Power Consumption Reduction, Noise Power -65 dBm}
\resizebox{0.49\textwidth}{!}{ 
\begin{tabular}{|c|c|c|c|c|c|c|c|c|c|c|}
\hline
\textbf{Time Sharing} & \multicolumn{3}{|c|}{\textbf{Transmit}} & \multicolumn{3}{|c|}{\textbf{Data Rates}} & \textbf{Time-shared} & \multicolumn{3}{|c|}{\textbf{SIC Decoding}} \\ 
 &  \multicolumn{3}{|c|}{\textbf{Power (dBm)}} &  \multicolumn{3}{|c|}{\textbf{(Mbps)}} & \textbf{Fraction} & \multicolumn{3}{|c|}{\textbf{Order}} \\ 
\hline
\multirow{3}{*}{\checkmark} & \multirow{3}{*}{15} & \multirow{3}{*}{14.3} & \multirow{3}{*}{15} & 398.01 & 470.48 & 632.23 & 0.52 & 3 & 2 & 1 \\
& &  &  & 691.78 & 242.32 & 565.91 & 0.17 & 1 & 3 & 2 \\
& &  &  & 565.91 & 691.78 & 242.32 & 0.31 & 2 & 1 & 3 \\ \hline 
\ding{55} & 15.8 & 16 & 15.4 & 500 & 500 & 500 & 1.00 & 3 & 2 & 1 \\
\hline
\end{tabular}
}\label{tab:time-sharing2}
\end{table}

\begin{table}[t!]
    \caption{2 Small Experiments showing Lower Power Consumption of Proposed Algorithm compared to OMA}
    \centering
    \resizebox{8.5cm}{!}{
    \begin{tabular}{|c|c|c|}
    \hline 
    \textbf{Parameter}& \textbf{Case 1}& \textbf{Case 2}\\
    \hline 
    \textbf{Antennas/user}& 1 & 1 \\
    \textbf{AP Antennas}& 2 & 2\\
    \textbf{Users}& 2 & 3 \\
    \textbf{Distance from AP (m)}& \{3, 3\} & \{3, 3, 3\} \\
    \hline
    \textbf{Data Rates (OMA) (Mbps)}& \{495.0, 495.0\} & \{454.3, 440.0, 442.1\} \\
    \textbf{Data Rates (Proposed) (Mbps)}& \{495.0, 495.0\} & \{454.3, 440.0, 442.1\}\\
    \hline
    \textbf{Transmit Power (OMA) (dBm)}& \{15, 15\} & \{15, 15, 15\} \\
    \textbf{Transmit Power (Proposed) (dBm)}& \{10.8, 15.2\} & \{10.9, 9.8, 9.9\} \\
    \hline 
    \end{tabular}}
    \label{tab:scearios-single}
\end{table}

\begin{table}[t]
\caption{Transmit Power Consumption versus Number of AP Antennas for Single Antenna Per User, $R = 200 \textrm{ Mbps}$, $U = 3$, $L_{xu} = 1$, $\textrm{distance from AP} = \{3m,3m,3m\}$, transmit power values are relative to 15 dBm}
\centering
\resizebox{9cm}{!}{
\begin{tabular}{|c|c|c|c|c|}
\hline
\multirow{2}{*}{\begin{tabular}[c]{@{}c@{}}\textbf{AP} \\ \textbf{Antennas}\end{tabular}} & \multicolumn{2}{c|}{\textbf{Linear Receiver}} & \multicolumn{2}{c|}{\textbf{Proposed Algorithm}} \\ \cline{2-5} 
                     & \textbf{Transmit Power}& \textbf{$\textrm{Power}_{\textrm{avg}}$}& \textbf{Power}& \textbf{$\textrm{Power}_{\textrm{avg}}$}\\ \hline
1                    & [1,  1,  1] & 1& [0.5721,    0.5417,    1.1325]& 0.7488\\ \hline
2                    & [0.9, 0.9,  0.9] & 0.9& [0.1354,    0.0422,    0.1552]& 0.1109\\ \hline
3                    & [0.75, 0.75, 0.75] & 0.75& [0.0294,    0.0750,    0.0893]& 0.0646\\ \hline
4                    & [0.3, 0.3, 0.3] & 0.3& [0.0214,    0.0508,    0.0556]& 0.0426\\ \hline
\end{tabular}}
\label{tab:energy-compare}
\end{table}

\begin{table}[t]
\caption{Transmit Power versus Number of AP Antennas for Single Antenna Per User, Fixed energy for linear receiver, $U = 3$, $L_{xu} = 1$,  $L_{y}=2$, $\textrm{distance from AP} = \{3m,3m,3m\}$, Transmit Power values are relative to 15 dBm}
\centering
\resizebox{9cm}{!}{
\begin{tabular}{|c|c|c|c|c|}
\hline
\multirow{2}{*}{\begin{tabular}[c]{@{}c@{}}\textbf{AP} \\ \textbf{Antennas}\end{tabular}} & \multicolumn{2}{c|}{\textbf{OMA}} & \multicolumn{2}{c|}{\textbf{Proposed Algorithm}} \\ \cline{2-5} 
                     & \textbf{Transmit Power}& \textbf{Data Rates}& \textbf{Transmit Power}& \textbf{Data Rates}\\ \hline
1                    & [1,  1,  1] & [199,	119, 236]& [0.7745, 0.2652, 1.1374]& [199,	119, 236]\\ \hline
2                    & [1,  1,  1] & [172, 147, 156]& [0.1187,    0.1989,    0.2021]& [172, 147, 156]\\ \hline
3                    & [1,  1,  1] & [227, 202, 220]& [0.1453,  0.1132,  0.1370]& [227, 202, 220]\\ \hline
4                    & [1,  1,  1] & [353, 291, 332]& [0.2920, 0.2696, 0.2853]& [353, 291, 332]\\ \hline
\end{tabular}}
\label{tab:energy-compare2}
\end{table}

\begin{figure}[t]
    \centering
    \includegraphics[trim = {0, 180, 0, 180}, clip, height = 4.5cm]{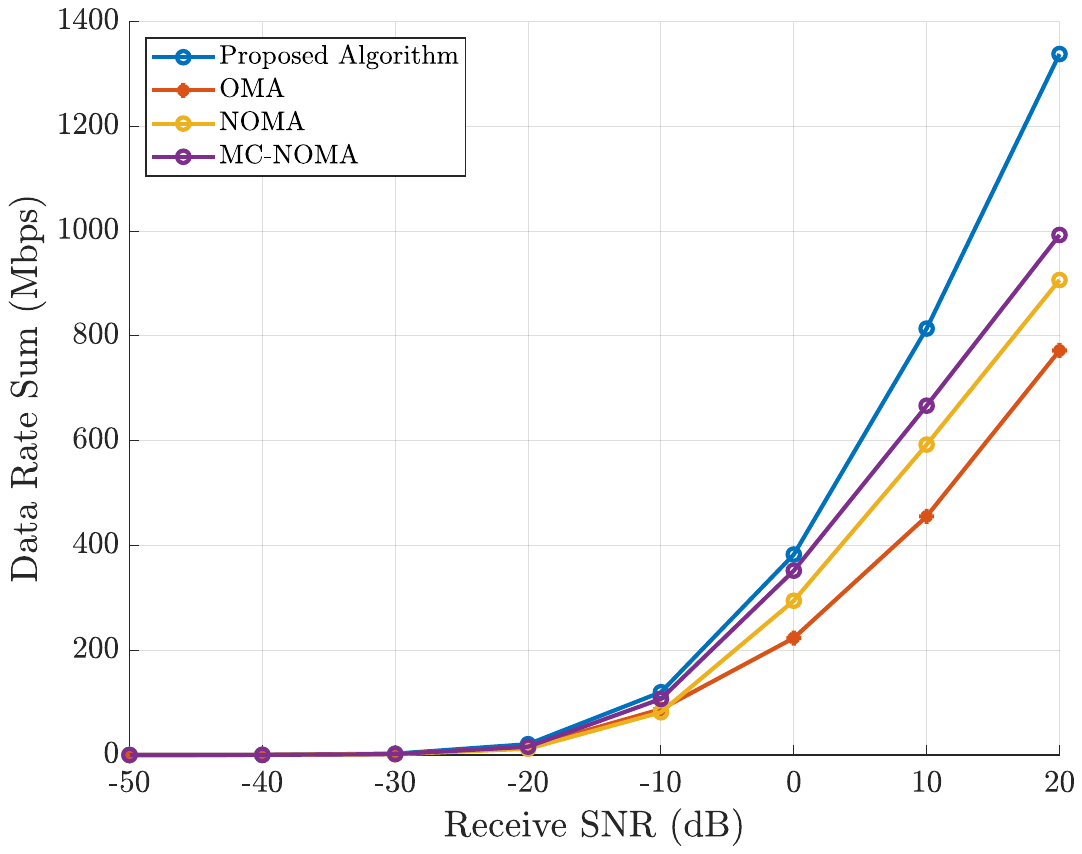}
    \caption{Sum Rate versus receive SNR for Single Antenna Per User with distances from AP=\{3m, 3m, 3m\}: Baselines and Proposed Algorithm}
    \label{fig:data-rate-snr-single}
\end{figure}




\begin{figure}
    \centering
    \includegraphics[trim = {0, 180, 0, 180}, clip, height = 4.5cm]{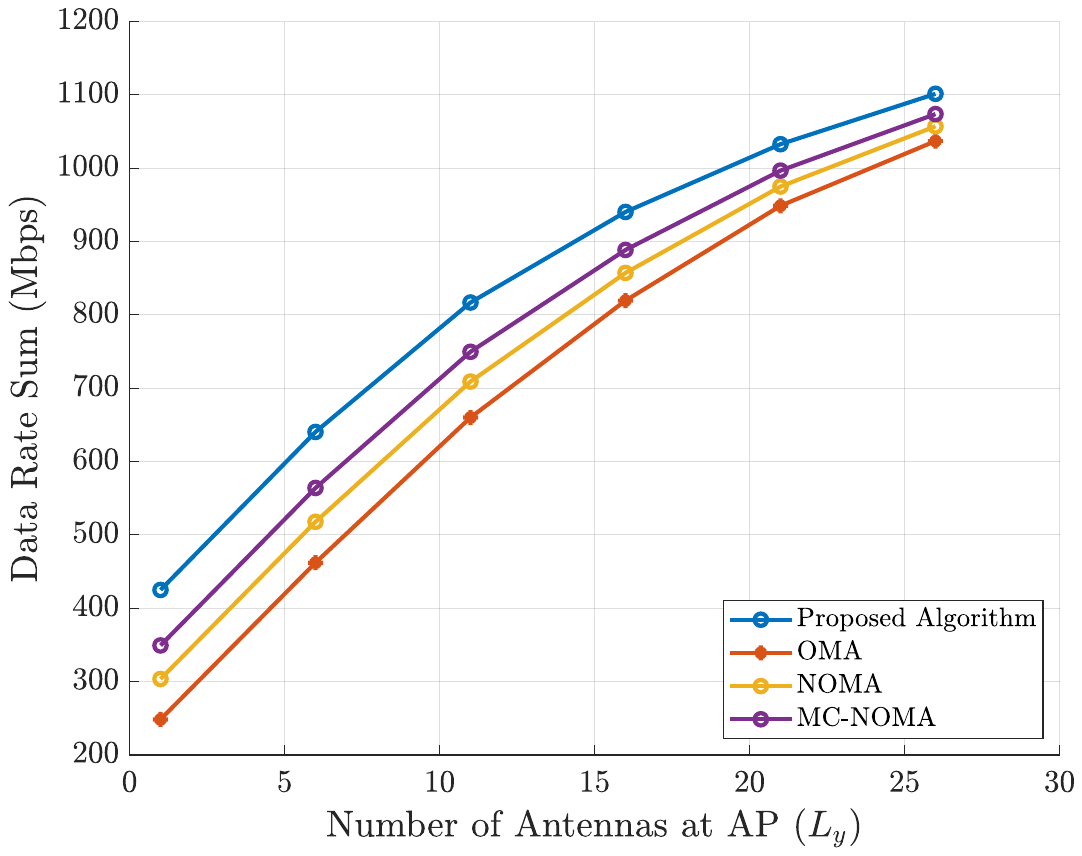}
    \caption{Sum Rate versus Number of AP Antennas, 0 dB Receive SNR at AP}
    \label{fig:vary-AP-antennas}
\end{figure}

\begin{figure}
    \centering
    \includegraphics[trim = {0, 180, 0, 180}, clip, height = 4.5cm]{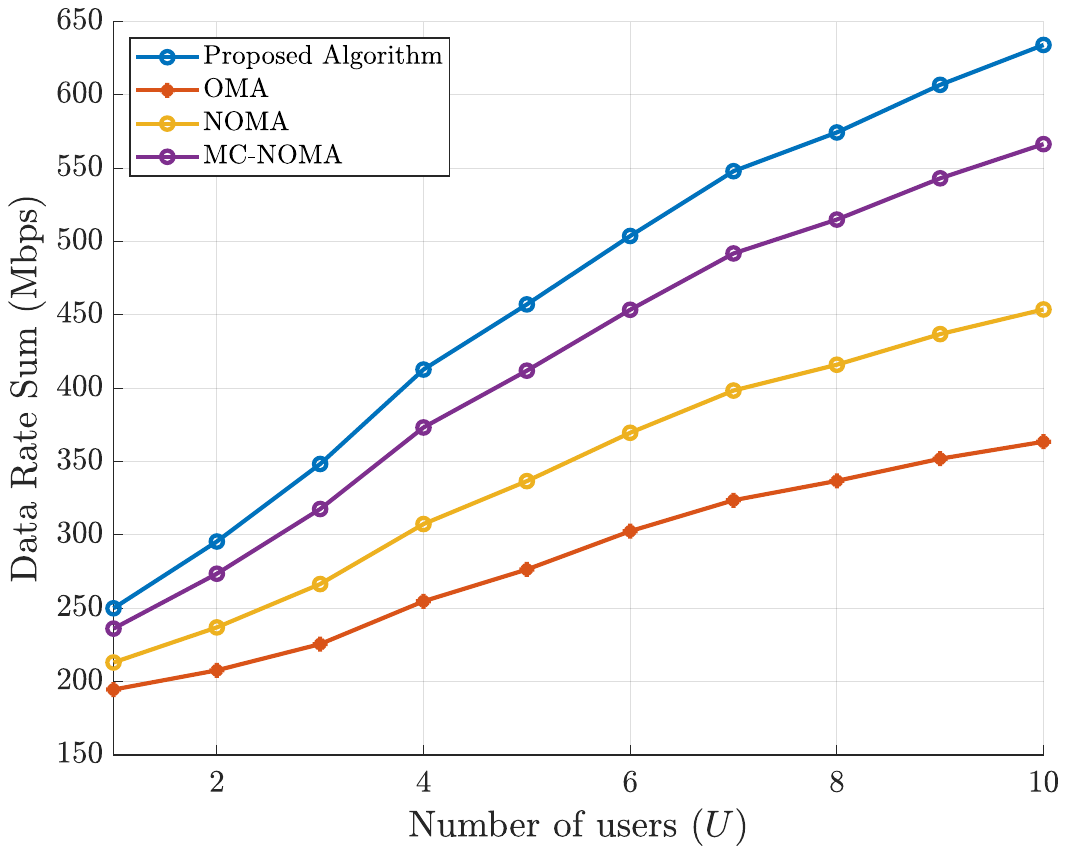}
    \caption{Sum Rate versus Number of Users, 0 dB Receive SNR at AP}
    \label{fig:vary-users}
\end{figure}

\begin{figure}
     \centering
     \hspace{-0.8cm}
     \begin{subfigure}{0.2\textwidth}
         \centering
         \includegraphics[trim = {40, 180, 50, 200}, clip, height = 3.5cm]{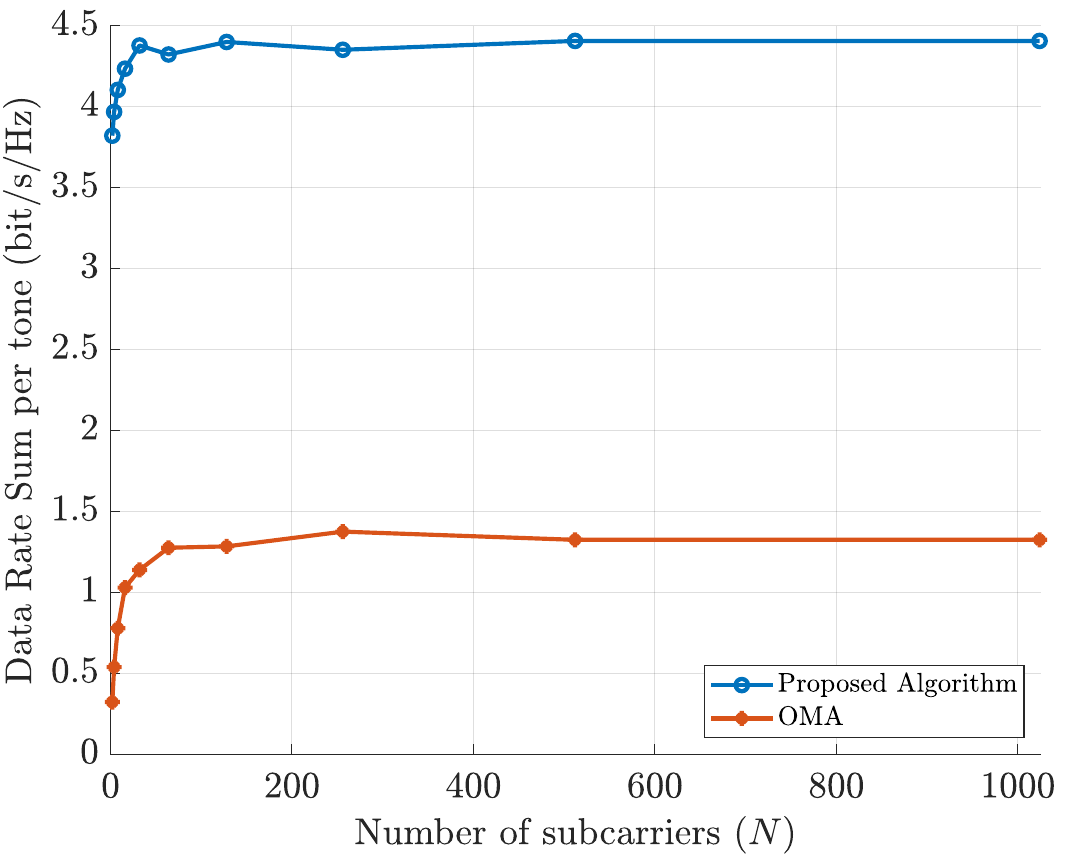}
         \caption{Spectral efficiency of three users with distance from AP = $\{3m, 3m, 3m\}$}
         \label{fig:333}
     \end{subfigure}
     \hspace{0.8cm}
     \begin{subfigure}{0.2\textwidth}
         \centering
         \includegraphics[trim = {40, 180, 50, 200}, clip, height = 3.5cm]{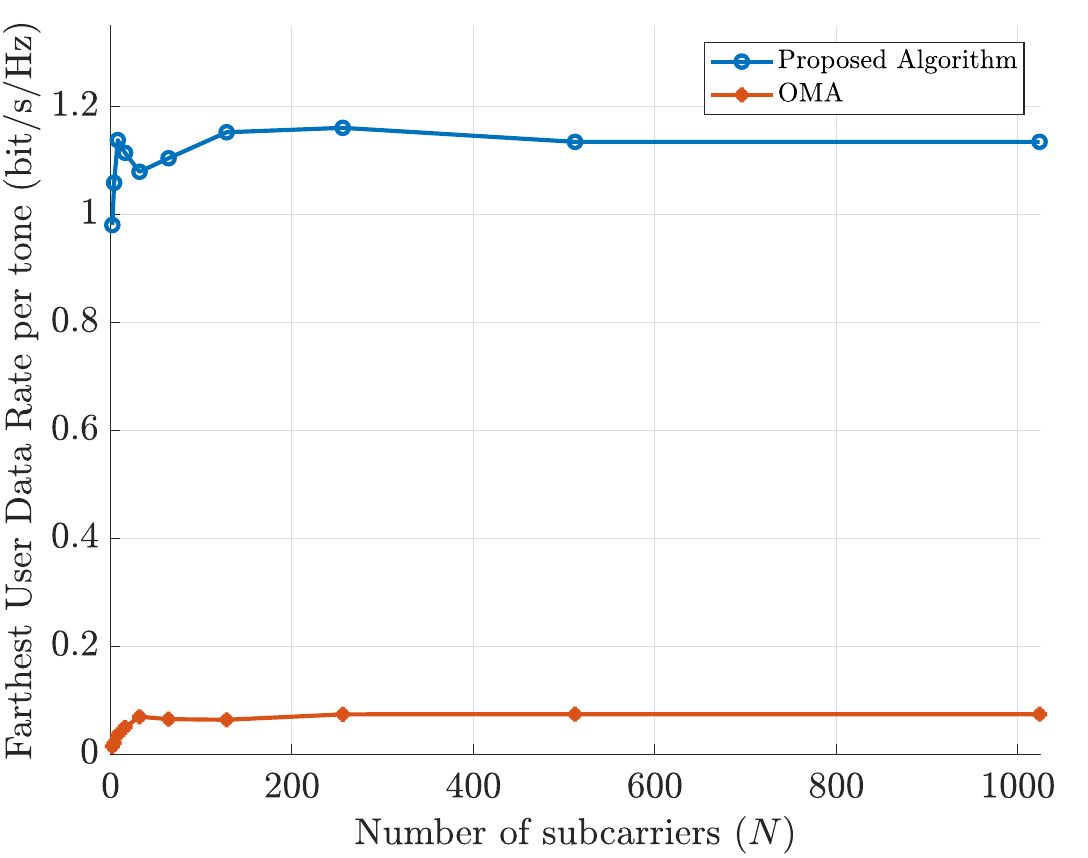}
         \caption{Spectral efficiency of farthest user with distance from AP = $\{3m, 5m, 8m\}$}
         \label{fig:358}
     \end{subfigure}
     \caption{Spectral efficiency sum versus number of subcarriers, single antenna per user, and two antennas at AP}
\end{figure}


The experimental setup comprises a distributed wireless VR gaming environment involving \( U \) users, each of which contains 1 antenna. For this task, each of the users requires data rates of 500 Mbps or higher in the uplink. \( L_{xu} \) is an $1\times  U$ array, that represents the number of antennas at each user's device. Additionally, the AP has \( L_y \) antennas. The users' spatial positioning relative to the AP has a uniform random distribution, with distances between \( d_{\text{min}} \) and \( d_{\text{max}} \). The channel impulse responses between user devices and the AP are sampled uniform randomly from a dataset of over 10,000 experimentally collected Wi-Fi channel realizations. These realizations use a Wi-Fi 802.11ax model \cite{802.11ax}, specifically tailored for indoor residential (model B) and indoor office (model D). The experiment randomly selects one channel impulse response for each pairing of the user device antenna and AP antenna. The simulations multiply additional components to this channel impulse response corresponding to path loss, shadow fading and Rayleigh fading. 

The path loss and shadow fading loss are:
\begin{equation}
    \begin{aligned}
        L(d) &= L_{path}(d) + L_{shadow}(d) \quad \textrm{dB}  \qquad \qquad d \leq d_{break} \\
    &= L_{path}(d_{break}) + L_{shadow}(d_{break}) + \\ & \qquad \qquad \qquad 35 \: \operatorname{log}_{10}(\frac{d}{d_{break}}) \quad \textrm{dB}  \qquad d>d_{break}
    \end{aligned}
\end{equation}
where the break-point distance is $d_{break} = 5\:m$, and 
\begin{align}
    L_{path} =  20 \: \operatorname{log}_{10}(f) + 20\: \operatorname{log}_{10}(d)  - 147.5 \textrm{ dB}
\end{align}
$L_{shadow}$ is a random log-normal variable sampled from $\frac{1}{\sqrt{2 \pi \sigma^{2}_{z}}}\exp\left({-L_{shadow}(d)^{2}/(2\sigma^{2}_{z})}\right)$, 
$\sigma_{z}$ is $3 \textrm{ dB}$ before the breakpoint distance and $4 \textrm{ dB}$ after the breakpoint. The center frequency is, $f_{c}=5\textrm{ GHz}$ and the number of OFDM subcarriers $N$ varies. Table~\ref{tab:experiment-parameters} shows the parameters used for the experiments. We vary the transmit SNR values and assess the resultant data rates achieved by the baselines and the proposed algorithm. 

Table \ref{tab:scearios-single} presents two experiments with 2 and 3 participating users respectively. The transmit power for each of the users is constant at 15 dBm for the OMA baseline, following WiFi standards. We then compute the power required to achieve the same data rates as the OMA baseline, using the proposed algorithm. The average power consumption for achieving the same data rates using the proposed algorithm is 35\% lower than OMA. Column 1, which represents a full rank channel with 2 users and 2 AP antennas, still shows significant energy savings of 21\% with the proposed algorithm. This demonstrates proposed algorithm's use of cross-talk among the user channels to achieve the same data rates under lower transmit power. Column 2 represents a low rank channel scenario, with 3 users and 2 AP antennas. Here we see a more pronounced effect and higher energy savings of 46.3\% using the proposed algorithm as compared to OMA. This shows that the proposed algorithm, equipped with optimal decoding order derivation and time sharing, achieves significantly better performance than linear receivers, especially under low-rank channel scenarios.

Table~\ref{tab:energy-compare} demonstrates the proposed algorithm's superiority over OMA in power efficiency across various AP antenna counts, maintaining a fixed data rate of 200 Mbps/user. The proposed algorithm achieves the same data rates sum with $72.5$\% lower power consumption on average. The power savings become more pronounced as AP antennas decrease, attributed to the channel's lower rank in these scenarios. This further validates the proposed algorithm's ability to use the channel crosstalk and derive the optimal decoding order and time sharing in low rank channel scenarios. 
Table~\ref{tab:energy-compare2} demonstrates the proposed algorithm's transmit power values that achieve the same data rates as achieved by the OMA-based linear receiver, with increasing number of AP antennas. For this figure, the power used for OMA remains constant at 15 dBm per user. It plots the data rates achieved by the linear receiver under these constraints. Then the energies required by the proposed algorithm achieve the same data rates. The proposed algorithm achieves, on an average, $70.7$\% lower power consumption compared to linear receiver. The reduction in power consumption occurs because the proposed algorithm uses the crosstalk between signals of different users and derives the optimal decoding order and time sharing. Hence it can reach higher data rates compared to linear receiver.

Fig.~\ref{fig:data-rate-snr-single} shows the sum rate across 3 users located at equal distances of 3 m from the AP, as a function of receive SNR at the AP. The proposed algorithm outperforms the OMA, NOMA and MC-NOMA baselines by 39\%, 28\%, and 16\% respectively in terms of average sum data rates, at varying SNR levels. The sum rates achieved by the proposed algorithm increases at a significantly steeper slope compared to the baselines as well. The OMA baseline, which uses the linear receiver (as implemented in current WiFi standards) achieves the lowest sum rates because of its inability to utilize the channel cross talk. The cross talk is utilized by the NOMA, MC-NOMA baselines, as well as the proposed algorithm. However, the NOMA method allocates the entire available bandwidth to all pariticipating users, and does optimal power allocation based on that. The improved NC-NOMA, which does a power-subcarrier allocation similar to the proposed algorithm, still makes a heuristic assumption on the SIC decoding order based on the channel strength order. This heuristic decoding order turns out to be suboptimal for vector channels, which is the case here since there are 2 AP antennas. The optimum power allocation derivation and time sharing-based adaptive decoding order varying helps the proposed algorithm achieve significantly better performance.

Fig.~\ref{fig:vary-AP-antennas} shows the sum data rates achieved at 0 dB receive SNR at the AP, versus varying number of AP antennas. The proposed algorithm outperforms the OMA, NOMA and MC-NOMA baselines at all AP antenna values. As the number of AP antennas increases, the uplink MAC channel goes from low rank to full rank, i.e. there are sufficient degrees of freedom to support 3 users. Thus, we see that the advantage provided by the proposed algorithm is especially significant in low rank conditions, i.e., when the number of AP antennas is low. Fig.~\ref{fig:vary-users} shows the sum data rates achieved at 0 dB receive SNR at the AP, versus varying number of users. The proposed algorithm achieves higher sum rates compared to all baselines by a significant margin. As the number of users increases, the gap between the proposed algorithm and the baselines becomes more prominent. This is because, as the number of users increases, the low rank effect of the uplink MAC channel, with only 2 AP antennas, becomes more pronounced. The proposed algorithm provides the desired data rates even in extreme low rank channel scenarios, while the baselines fail to do so.

We also analyze the effect of varying the number of OFDM subcarriers $N$ on the achieved data rates. Fig.~\ref{fig:333} and Fig.~\ref{fig:358} compare the spectral efficiencies of the proposed algorithm against OMA baseline in users equidistant (3m, 3m, 3m) from the AP, and non-equidistant (3m, 5m, 8m) from the AP, respectively. We provide as an input to the proposed algorithm the required spectral efficiency values of 5 bits/s/Hz for each of the three participating users. This translates into data rate requirements of 400 Mbps/user for an 80 MHz channel bandwidth. Fig.~\ref{fig:333} shows the sum of the spectral efficiency values across all three users, as a function of $N$. As per Fig.~\ref{fig:333}, the proposed algorithm achieves sum rates of 4.41 bits/s/Hz with $N=1024$, outperforming the OMA baseline's 1.32 bits/s/Hz. It is interesting to analyze if the proposed algorithm is able to achieve these required data rates for each of the participating users, even when their distances from the AP are unequal. To do so, Fig.~\ref{fig:358} shows the spectral efficiency achieved by the farthest user in a (3m, 5m, 8m) setting. The farthest user at 8m from the AP has the worst channel conditions. In Fig.~\ref{fig:358}, the proposed algorithm ensures even the farthest user achieves the required data rates of 500 Mbps, despite having the worst channel with the highest path losses, leveraging optimum power allocation and time sharing. The linear receiver in the OMA baseline fails to achieve this, and hence the farthest user saturates at 0.07 bits/s/Hz. This demonstrates the proposed solution's superiority in energy and data rate efficiency, especially in low-rank channel conditions, making it a promising alternative to current OMA and NOMA methods.


\section{Conclusions and Future Directions}

The proposed optimum power-subcarrier allocation and time-sharing algorithm for multi-user uplink WiFi channels significantly enhances data rates and energy efficiency compared to OMA, NOMA and MC-NOMA baseline methods. The data rate gains become significant especially in low-rank wireless channel scenarios. Future directions include the integration of machine learning models to optimize power allocation dynamically, catering to real-time changes in channel conditions.
\section{Acknowledgement}
This research was funded through support and collaboration from Stanford, Ericsson, Intel, and Samsung.

\footnotesize
\bibliographystyle{IEEEtran}
\bibliography{main.bib}

\end{document}